# Transient superionic state in ultrafast-irradiated post-transition metal oxides

N. Medvedev[1,2,*], N. Nikishev[1,3], A. Artímez Peña[1,3]

*1) Institute of Physics, Czech Academy of Sciences, Na Slovance 1999/2, 182 00 Prague 8, Czech Republic*
*2) Institute of Plasma Physics, Czech Academy of Sciences, Za Slovankou 3, 182 00 Prague 8, Czech Republic*
*3) Czech Technical University in Prague, Břehová 7, 115 19 Prague 1, Czech Republic*

## Abstract

Matter under irradiation may enter unusual transient states, outside of its equilibrium phase diagram. One of such states is a superionic-like state, in which one sublattice of a compound liquifies, whereas another one remains solid. Here, we study theoretically post-transition metal oxides under ultrafast excitation of its electronic system, identifying which compounds produce such a superionic state. It is shown that oxides with sufficiently sparce metallic sublattices (e.g. corundum structure) generally form transient superionic states via nonthermal phase transition. More closely packed lattices (such as the zinc-blend structure in ZnO and CdO) do not exhibit superionicity. Tl and Pb oxides only enter thermally-produced superionic states (induced by the atomic heating via electron-phonon coupling), but not nonthermal ones. Sn and Bi oxides demonstrate states that cannot be clearly classified, in which oxygen subsystem diffuses significantly more and faster than the metallic one, but the metallic one is not stable as it would be in a truly superionic state.

## I. Introduction

Ultrafast irradiation of materials with photons or charged particles induces nonequilibrium kinetics which may create unusual transient states of matter [1–7]. Those states outside of the phase diagram evolve quickly, eventually relaxing towards thermodynamic equilibrium [8–13]. Under some conditions, they may end up in new phases of materials, unachievable by other means [2–4,8,14].

Recently, it has been predicted that certain oxides and polymers upon ultrafast excitation of their electronic system transiently produce superionic states [15–18]. Superionic states are characterized by simultaneous solid and liquid properties in a material: in a compound, one (or more) atomic subsystem is in a crystalline lattice, whereas another one is fluid [19,20]. Superionic states are widely used in solid-state lithium batteries, where Li ions are diffusing in a liquid-like state, embedded into a solid structure of other elements [21,22]. Superionic water or ice is predicted to exist in giant icy planets, where hydrogen is expected to be in a liquid state within a solid oxygen lattice [20,23]. These facts trigger practical as well as fundamental interest in the superionic states insufficiently studied so far.

Transient superionic state may be produced in some material under irradiation within a range of deposited doses (energy densities), at which one of the sublattices "melts", whereas the other remain relatively stable [15–18]. That is a consequence of a different strength of interatomic bonds for different elements in a compound, and different modification of those bonds, selectively rupturing for one element but not another. To date, there is no predictive theory to explain which materials produce superionic state, and which do not. The radiation physics and chemistry community is currently at the stage of accumulation of knowledge, to be able to discern some patterns and deduce some general conclusions.

[*] Corresponding author: nikita.medvedev@fzu.cz; ORCID: 0000-0003-0491-1090

Contributing to this search, the present work studies theoretically post-transition metal oxides under ultrafast irradiation. We analyze 12 different materials, from aluminum to bismuth oxide, identifying which of the compounds form transient superionic states, and at which conditions. We identify the damage threshold doses and mechanisms taking place at such excitation levels. Interestingly, the results reveal that there may be different kinetic pathways to enter superionic states, characterized by either electronic excitation (nonthermal transition), or by atomic heating without the electronic excitation (thermal transition). Some materials exhibit a state that cannot be clearly classified as superionic or not, suggesting much richer landscape of possibilities than previously thought.

## II. Model

A state-of-the-art multiscale code, XTANT-3, is applied here to simulate post-transition metal oxides under ultrafast laser irradiation [24]. The code combines a few models into one interconnected framework with feedbacks. All the details of the models and numerical technics used are thoroughly described in the code's manual [24]. Here, we briefly outline the essentials of the models and list the parameters applied in the current work.

The photoabsorption, excited electron kinetics, and Auger decays of core holes, are simulated with the transport Monte Carlo (MC) method. The photoabsorption cross sections for each shell of each element, as well as the ionization potentials and Auger decay times, are extracted from EPICS2025 database [25]. The electron impact ionization cross sections are calculated with the linear response theory within the single-pole approximation [26]. The (quasi) elastic scattering of fast electrons on atoms is modelled with the screened Rutherford cross section with the modified Molier screening parameter [27]. All electrons are traced with event-by-event MC simulation until they lose the energy below 15 eV threshold counted from the bottom of the conduction band. Auger cascades of core holes are traced until a hole transfers into the valence band. Electrons falling below the cutoff energy, and holes in the valence band, join the low-energy electron domain.

Low-energy electrons are traced with the Boltzmann equation (BE). In the homogeneous simulation box, it is reduced to the Boltzmann collision integrals, describing the electron-electron and electron-ion (electron-phonon) scattering. The electron-electron interaction is treated within the relaxation time approximation, which in the limit of the instantaneous thermalization produces the Fermi-Dirac distribution [10]. The electron-phonon scattering is calculated with the nonperturbative dynamical coupling approach [28,29]. The distribution function describes the evolution of the electron fractional population numbers on the transient energy levels.

The evolution of the electron energy levels (band structure), as well as the interatomic potential, is traced with the transferable tight binding (TB) method [30]. Here, matsci-0-3 tight binding parameterization is used for $Al_2O_3$ [31], whereas PTBP parametrization is applied for all other considered compounds [32]. Diagonalization of the Hamiltonian on each timestep of the simulation provides us with the electronic energy levels, whereas the derivative of the potential energy (with the transient electron populations) form the interatomic forces. Due to evolution of the electronic populations (BE above), this method is capable of describing the effects associated with the changes of the interatomic potential induced by the electronic excitation, such as nonthermal melting [1,33].

The atomic dynamics is simulated using molecular dynamics (MD) method with the interatomic forces calculated from the TB. Martyna-Tuckerman fourth order algorithm is used for propagating atomic trajectories, with the typical timestep of 0.2-1 fs, depending on the particular material, ensuring stability and energy conservation. We apply periodic boundary conditions, mimicking bulk material with the microcanonical (NVE) ensemble. The energy provided by the nonadiabatic electron-ion (electron-phonon) coupling and by quasi-elastic scattering of high-energy electrons is delivered to atoms on each timestep *via* the velocity scaling algorithm [24].

The supercells contained the following numbers of atoms: 240 for $Al_2O_3$; 360 for α-phase of $Ga_2O_3$; 320 for $In_2O_3$, $Tl_2O_3$, $Pb_2O_3$, and β-phase of $Ga_2O_3$; 256 for SnO; 288 for $Tl_2O$; 256 for PbO; 160 for $Bi_2O_3$ and $BiO_2$; and 216 for CdO and ZnO. The oxides $Al_2O_3$, $Ga_2O_3$, $In_2O_3$, $Tl_2O_3$, $Pb_2O_3$, and $Bi_2O_3$ are simulated in the corundum structure, whereas the others are in their respective most stable allotropes [34]. Convergence with respect to the number of atoms in XTANT-3 simulations was previously shown to be achieved for ~200 atoms [28].

In a simulation, photon energy of 92 eV (a typical energy of FLASH free-electron laser [35]), pulse duration of 10 fs (full width at half maximum of a Gaussian pulse centered at 0 fs), and various deposited doses (energy densities) are applied. XTANT-3 was previously validated in simulations of various materials, providing reasonable agreement with experiments [11,36,37].

## III. Results and discussion

We perform a series of simulations for each material to identify the threshold dose at which the material damage onsets. The state produced is identified by the element-resolved mean displacement of atoms: superionic state is characterized by continuous increase of the displacements of one subsystem (diffusion, liquid state) and saturated displacement of the other subsystem (indicating a solid structure) [15,17]. In contrast, melting is identified by the diffusion (continuous increase of the displacement) in both subsystems.

The calculated damage thresholds and their types are summarized in Table 1. The results qualitatively agree with previous reports on the few materials: the superionic state in $Al_2O_3$ was predicted in Refs. [15,17]; the superionic state in $Ga_2O_3$ and $In_2O_3$ (and its absence in $In_2S_3$) were predicted by means of DFT calculations within the Born-Oppenheimer approximation (no electron-phonon coupling) [16].

Our results agree with those DFT calculations, thus validating XTANT-3 modeling: oxygen subsystem in these oxides turns liquid upon electronic excitation, whereas the metallic one remains crystalline. XTANT-3 simulations demonstrate that the presence of the nonadiabatic electron-phonon coupling does not preclude the formation of superionic states in these materials, however, it makes this state transient. Once the atomic system is heated by the electrons, the metallic subsystem also melts, see Figure 1.

*Table 1. Damage threshold and transition types induced in various materials.*

| Material | Transition type | Threshold dose (eV/atom) |
|---|---|---|
| $Al_2O_3$ | Nonthermal superionic | 2.5 |
| α-$Ga_2O_3$ | Nonthermal superionic | 0.25 |
| β-$Ga_2O_3$ | Thermal superionic | 0.2 |
| $In_2O_3$ | Thermal superionic | 0.4 |
| SnO | Nonthermal superionic liquid | 0.3 |
| $Tl_2O_3$ | Nonthermal superionic | 0.6 |
| $Tl_2O$ | Thermal melting | 0.3 |
| $Pb_2O_3$ | Nonthermal superionic | 0.05 |
| PbO | Thermal superionic | 0.2 |
| $Bi_2O_3$ | Nonthermal superionic liquid | 0.1 |
| $BiO_2$ | Nonthermal superionic liquid | 0.05 |
| ZnO | Nonthermal melting | 0.4 |
| CdO | Nonthermal melting | 1.5 |

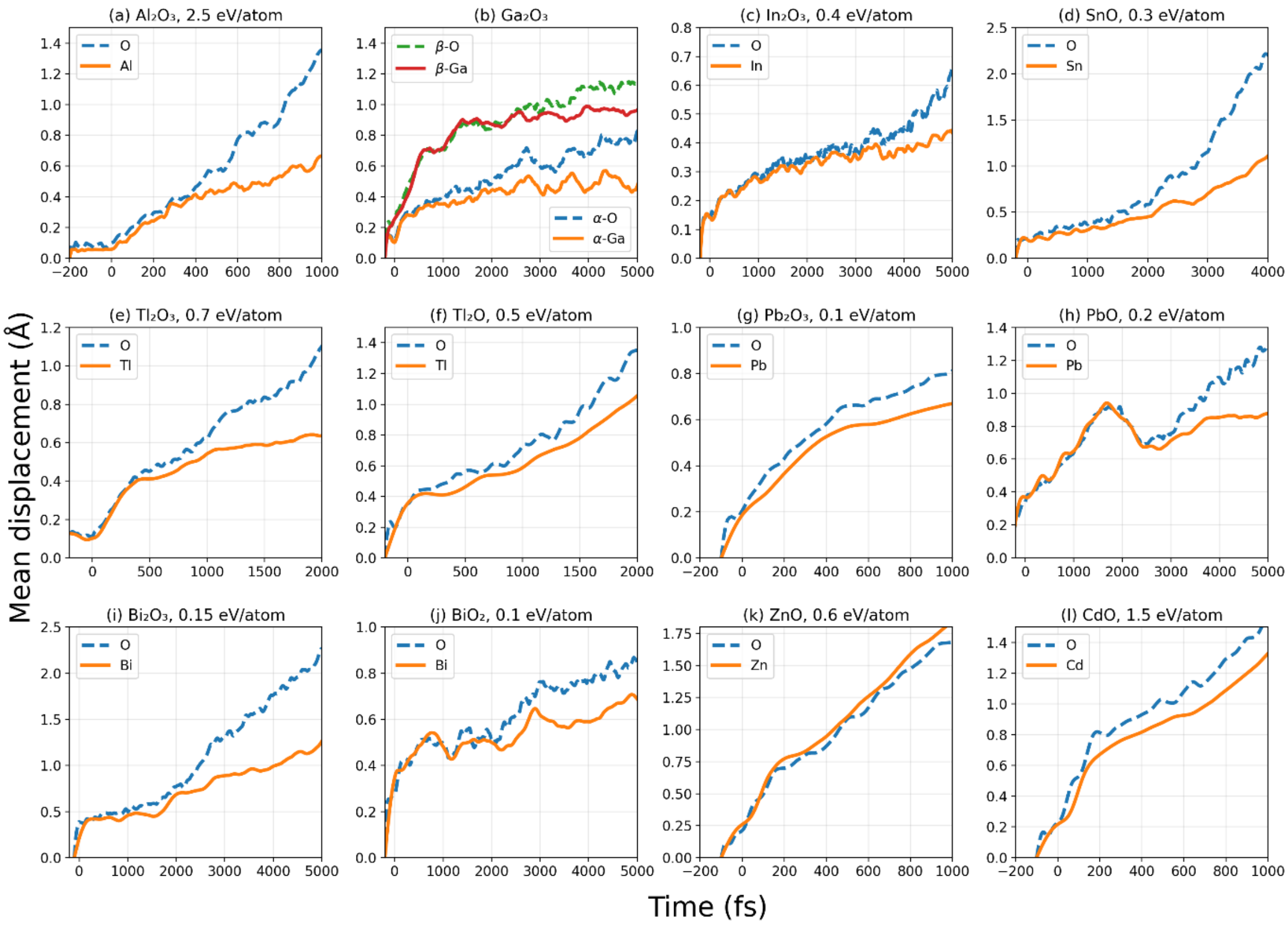

*Figure 1. Mean displacement in various materials under irradiation with above-threshold dose. (a) $Al_2O_3$, 2.5 eV/atom. (b) $Ga_2O_3$, 2.5 eV/atom in α phase, and 2.0 eV/atom in β phase. (c) $In_2O_3$, 2 eV/atom. (d) SnO, 0.3 eV/atom. (e) $Tl_2O_3$, 0.7 eV/atom. (f) $Tl_2O$, 0.5 eV/atom. (g) $Pb_2O_3$, 0.1 eV/atom. (h) PbO, 0.2 eV/atom. (i) $Bi_2O_3$, 0.15 eV/atom. (j) $BiO_2$, 0.1 eV/atom. (k) ZnO, 0.6 eV/atom. (l) CdO, 1.5 eV/atom.*

In $Al_2O_3$, α-$Ga_2O_3$, $Tl_2O_3$, and $Pb_2O_3$, the superionic state forms at sub-picosecond timescales, indicating that it is a nonthermal process (again, agreeing with the conclusions of Ref. [16]). Oxygen disordering takes place due to electronic excitation (high electronic temperature), which modifies the interatomic potential.

In contrast, $In_2O_3$, PbO, and β-$Ga_2O_3$ turn superionic at longer timescale, when electronic system is already (almost) relaxed, and atomic temperature is close to its maximum, see Figure 2 (notice atomic temperatures reaching a plateau by the time of onset of the superionic states). This indicates that the superionic state is formed *via* thermal process: it is a precursor to complete thermal melting. During this stage, one subsystem (oxygen) melts first, while the metallic one remains stable.

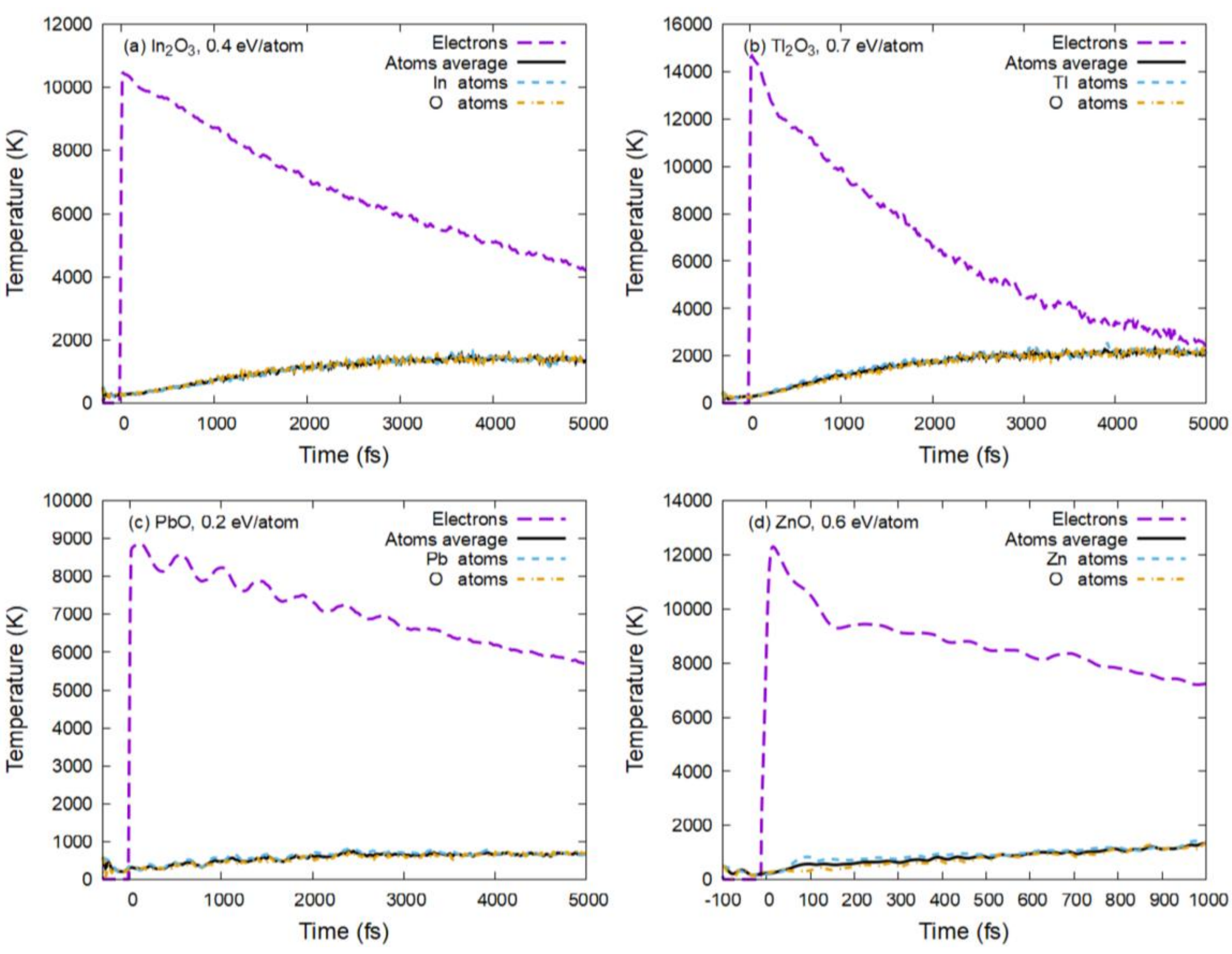

*Figure 2. Electronic and atomic (average and element-specific) temperatures in (a) $In_2O_3$ in irradiated with the dose 2 eV/atom; (b) $Tl_2O_3$ irradiated with 0.7 eV/atom; (c) PbO irradiated with 0.2 eV/atom; (d) ZnO irradiated with 0.6 eV/atom.*

Figure 2 shows that the partial temperatures of different elements are almost the same during the entire phase transition and after it in all cases, except for ZnO where Zn temperature slightly overshoots oxygen temperature just after the energy deposition, at ~50 fs. This selective acceleration of one species indicates that they are more sensitive to modification of the interatomic potential induced by the electronic excitation, as was previously reported for some materials [36,38]. Their temperatures equilibrate after ~500 fs.

We found no superionic state produced in irradiated $Tl_2O$, ZnO, and CdO. The first one melts thermally due to atomic heating via electron-phonon coupling, whereas the latter two melt nonthermally at subpicosecond timescale. This can be seen by the ever-increasing atomic displacements at the same rates for both subsystems – oxygen and metallic.

Finally, SnO, $Bi_2O_3$, and $BiO_2$, exhibit a state that cannot be clearly classified as superionic: oxygen subsystem disorders much faster than the metallic one, which appears to be more stable, but not truly solid. Such a transient "superionic liquid" or a "defective superionic" state leads to a complete disorder at different rates for the different subsystems. That shows a very rich variety of transient states produced by ultrafast irradiation in various materials.

Having established the threshold doses, we may estimate the damage threshold fluences in the studied materials to help guiding future experiments. Under the assumption of the normal incidence and negligible particle and energy transport (including emission from the surface), the threshold fluence may be approximated as $F = D\lambda n_{at}$, where $D$ is the threshold dose, $\lambda$ is the photon attenuation length, and $n_{at}$ is the atomic concentration. The threshold fluences, corresponding to the doses from Table 1, are shown in Figure 3.

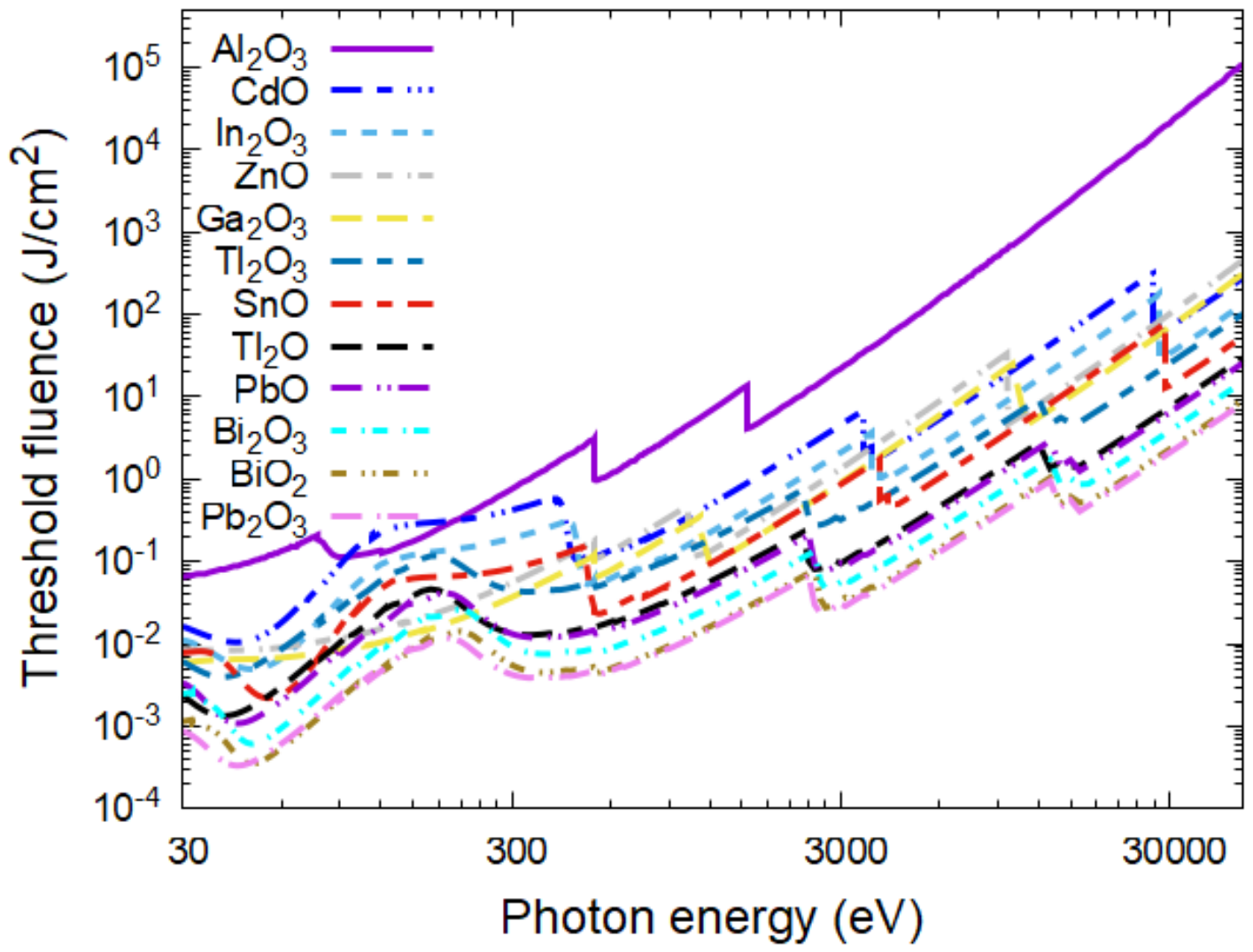

*Figure 3. Damage threshold fluences vs. incident photon energy in various materials.*

Our results contribute to the knowledge base of the produced transient superionic state upon irradiation exciting the electronic system. Although it is insufficient to deduce a theory or even a general pattern, under which conditions such states form, however, we can draw some conclusions from the comparative analysis of the materials studied. We notice that the metallic sublattice requires to be sparce to form superionic-like state *via* nonthermal phase transition: corundum structure is generally more prone to formation of superionic state than more close-packed structures.

We emphasize that this is only a correlation, but not a necessary, nor a sufficient condition: e.g., $In_2O_3$ in the corundum structure forms a superionic state *via* a different kinetic pathway (thermal vs. nonthermal in other oxides studied); and so does PbO, despite having a different crystalline structure. Counterexamples are $Fe_2O_3$ (not shown as it does not belong to the group of post-transition metal oxides) and $In_2S_3$ studied in [16], which do not appear to form a superionic state in a corundum structure – however, in a non-magnetic tight-binding-based simulation; it requires a separate dedicated study to validate our conclusion with a more advanced method.

We also note that the possibility of stabilizing such various transient states remains an open question. For example, it was predicted in Ref. [15] that the superionic state in $Al_2O_3$ may be stabilized by a high pressure. High temperatures, pressures, or other conditions may be explored in future research for the materials studied here, but this is beyond the scope of the present work.

## IV. Conclusions

In this work, we modelled the ultrafast response of post-transition metal oxides to intense ultrafast irradiation. Most oxides in corundum structure transiently enter a superionic-like state, characterized by a liquid-like oxygen subsystem coexisting with a solid metallic framework. We find that the metallic sublattice must be sufficiently sparce for this nonthermal transition to occur. In contrast, more closely packed zinc-blend structures of ZnO and CdO do not exhibit superionic states behavior under comparable excitation.

Tl- and Pb-based oxides display a different pathway: they form thermally-produced superionic states at near-melting atomic temperatures reached through by electron-phonon coupling. The two studied examples of Sn- and Bi-oxides demonstrate an intermediate regime in which the oxygen subsystem diffusing significantly faster than the metallic one – representing a state between a true superionic phase and a liquid.

These results reveal nonequilibrium states outside the equilibrium phase diagram, suggesting that ultrafast excitation may enable access to such unusual transient states. Such states in the studied oxides may motivate further research on possibilities to stabilize or exploit such transient phases and create materials with novel and enhanced properties.

## V. Conflicts of interest

There are no conflicts to declare.

## VI. Data and code availability

The code XTANT-3 is available from [24]. The photon attenuation lengths in the studied materials, used to calculate the damage threshold fluence from the absorbed dose, are available from https://doi.org/10.5281/zenodo.19811711.

# VII. Acknowledgements

We thank L. Juha, J. Chalupsky, and M. Kopecky for helpful and motivating discussions. Computational resources were provided by the e-INFRA CZ project (ID:90254), supported by the Ministry of Education, Youth and Sports of the Czech Republic. The authors gratefully acknowledge the financial support from the European Commission Horizon MSCA-SE Project MAMBA [HORIZON-MSCA-SE-2022 GAN 101131245]. NM thanks the partial financial support from the Czech Ministry of Education, Youth, and Sports (grant nr. LM2023068). NN appreciates the Researchers at Risk Fellowship provided by the Czech Academy of Sciences.